\documentclass[nofootinbib,preprintnumbers,prd,twocolumn,showpacs,amsmath,amssymb]{revtex4-1}
\usepackage{bm}
\usepackage{graphicx}
\usepackage{hyperref}

\newcommand{\rmd}{\mathrm{d}}   
\newcommand{\rmi}{\mathrm{i}}   
\newcommand{\rmc}{\mathrm{c}}
\newcommand{\rms}{\mathrm{s}}

\newcommand{\matt}{\text{matt}}
\newcommand{\tot}{\text{tot}}
\newcommand{\Gf}{G_\text{F}}

\newcommand{\tmu}{\tilde{\mu}}

\newcommand{\ket}[1]{|#1\rangle}

\newcommand{\td}{\frac{\rmd}{\rmd t}}

\newcommand{\caO}{\mathcal{O}}

\newcommand{\sfH}{\mathsf{H}} 
\newcommand{\bfH}{\mathbf{H}} 

\newcommand{\vp}{\protect{\vec{p}}} 
\newcommand{\hv}{\protect{\hat{v}}}

\newcommand{\btq}{\tilde{\mathbf{q}}}
\newcommand{\btd}{\tilde{\mathbf{d}}}
\newcommand{\bfd}{\mathbf{d}}
\newcommand{\bfq}{\mathbf{q}}

\newcommand{\bfs}{\mathbf{s}}

\newcommand{\bfg}{\mathbf{g}}
\newcommand{\bfe}{\mathbf{e}}
\newcommand{\bsigma}{\bm{\sigma}}

\begin{document}

\title{Flavor Oscillation Modes In Dense Neutrino Media}

\author{Huaiyu Duan}
\email{duan@unm.edu}
\affiliation{Department of Physics \& Astronomy, 
University of New Mexico,
Albuquerque, NM 87131, USA}
\date{\today}

\begin{abstract}
We study two-flavor neutrino oscillations in homogeneous neutrino
gases in which neutrinos and anti-neutrinos are in nearly pure weak
interaction states initially. We find that
the monopole and dipole oscillation modes can trigger flavor
instabilities in the 
opposite neutrino mass hierarchies in a nearly isotropic neutrino gas. 
For a class of simple neutrino systems 
we are able to identify the normal modes of neutrino oscillations 
in the linear regime. Our results provide new insights into the
recently discovered
multi-azimuthal angle (MAA) instability of neutrino
oscillations in supernovae.
\end{abstract}
 
\maketitle

\section{Introduction}
The neutrino oscillation phenomenon provides the first piece of solid
evidence of the physics beyond the standard model of particle physics
(see, e.g.\ \cite{Beringer:1900zz, Strumia:2006db} for
reviews). Although neutrino 
oscillations in vacuum and  matter are generally well understood,
flavor oscillations in dense neutrino media, especially in a
core-collapse supernova, are still being intensely 
investigated, and new phenomena are continously being discovered,
mostly through numerical simulations
(e.g.\ \cite{Duan:2006jv, Duan:2007bt, Dasgupta:2008my, Duan:2007sh,
  Dasgupta:2009mg, Gava:2009pj, Friedland:2010sc, Duan:2010bf,
  Chakraborty:2011nf, Cherry:2012zw}; see \cite{Duan:2010bg} for a
recent but incomplete review).
The complexity of the flavor oscillations in a dense neutrino medium
lies in the nonlinear, neutrino-neutrino scattering potential which
can engender collective neutrino flavor transformation.

To make the problem of collective neutrino oscillations more tractable,
certain assumptions were universally adopted. For example, it was assumed that 
both the number fluxes (or, in terms of transport theory,
the 0th angular moment of the distribution function)
and flavor content of neutrinos were
homogeneous and isotropic in the early Universe (e.g.\ \cite{Kostelecky:1993yt,
  Kostelecky:1993dm, Abazajian:2002qx}). For core-collapse
supernovae it was assumed that 
both the number fluxes and flavor content of neutrinos were spherically
symmetric about the center of the neutron star as well as
axially symmetric about any radial direction from the neutron star
\cite{Duan:2006an}.

In an early study \cite{Raffelt:2007yz} it was shown that angle-dependent flavor
transformation can occur in a homogeneous, symmetric, bipolar neutrino
gas in which
neutrinos and anti-neutrinos have equal densities. It turns out that this
phenomenon is much more common than previous expected.
It was recently pointed out \cite{Raffelt:2013rqa}
that, even if the flavor content of 
supernova neutrinos is (approximately) axially symmetric on the
surface of the neutron star, it will no longer be so after the
so-called multi-azimuthal angle (MAA) instability of neutrino
oscillations has fully developed. 
This claim seems to be supported by the linear
stability analysis \cite{Raffelt:2013rqa} and some direct numerical
simulations \cite{Mirizzi:2013rla, Mirizzi:2013wda}. This finding
calls into question almost all the previous calculations in 
literature because the existing calculations on collective neutrino
oscillations in supernovae, including those in 
\cite{Raffelt:2013rqa, Mirizzi:2013rla, Mirizzi:2013wda}, assume the
flavor content of supernova neutrinos to be spherically symmetric
about the neutron star, which
is not true once the axial symmetry is broken.

The stability analysis of the linearized of equations of motion 
(e.g.\ \cite{Banerjee:2011fj,Sarikas:2012ad,Vaananen:2013qja}) seems
straightforward by itself. However, the
mathematical criteria for neutrino flavor instabilities that are
derived from this method are not necessarily  
physically transparent or intuitive \cite{Raffelt:2013isa}.
To provide some insight for the arise of the MAA instability
a simplistic two-beam model was proposed 
\cite{Raffelt:2013isa}. It was shown that in this model two 
collective oscillation modes, known as the symmetric and
anti-symmetric modes, exist, and they 
behave like pendulums when the other mode is absent.

In this paper we focus on homogeneous neutrino gases in which
angle-dependent neutrino flavor transformation can be studied in a
self-consistent way.
In Section~\ref{sec:bipolar} we revisit the single-energy, symmetric
bipolar systems which were previously studied in literature.
We identify the normal modes of these systems in the linear
regime. We show that the monopole and dipole modes are unstable in the
opposite neutrino mass hierarchies, and that all other multipole modes
are stable in the linear regime.
In Section~\ref{sec:discussions} we generalize our study to other
homogeneous and isotropic neutrino gases. We also discuss how to use
our results to gain new insights into the newly discovered MAA instability of
neutrino oscillations in supernovae.
In Section~\ref{sec:conclusions} we give our conclusions.

\section{Normal Oscillation Modes in Monochromatic, Symmetric, Bipolar
  Neutrino Gases%
\label{sec:bipolar}}
\subsection{Equations of Motion}
For two-flavor neutrino oscillations the flavor content of a neutrino or
anti-neutrino can be described by its
``flavor isospin'' \cite{Duan:2005cp}. 
(See Appendix \ref{sec:nfis} for a more detailed discussion of the
neutrino flavor isospin.) 
The flavor isospin of a
neutrino can be defined as 
$\bfs=\psi^\dagger (\bsigma/2) \psi$, and that of an anti-neutrino can be
defined as  
$\bfs = (\sigma_2 \psi)^\dagger (\bsigma/2) (\sigma_2\psi)$,
where $\psi$ is the flavor wavefunction of the neutrino or
anti-neutrino, and $\sigma_i$ ($i=1,2,3$) are the standard Pauli
matrices. The flavor isospins of a neutrino and an anti-neutrino of
the same energy $E$ are distinguished by their oscillation frequencies
$\omega=\pm \Delta m^2/2E$, where the plus and minus signs apply to
the neutrino and the anti-neutrino, respectively, and $\Delta m^2$ is
the neutrino mass-squared difference.

The flavor isospin of a neutrino or an anti-neutrino obeys an
equation of motion 
(EoM) which is similar to that of a magnetic dipole coupled to
both the external magnetic field and other dipoles. For a
homogeneous neutrino gas one has
\begin{align}
\td\bfs_{\omega,\hv} &= 
 - 2\sqrt{2}\Gf \bfs_{\omega,\hv}\times \sum_{\omega',\hv'} 
(1-\hv\cdot\hv') n_{\omega',\hv'} \bfs_{\omega',\hv'}
\nonumber\\
&\qquad + \bfs_{\omega,\hv}\times\omega \bfH_0, 
\label{eq:eom-s}
\end{align}
where $\bfs_{\omega,\hv}$ is the flavor isospin for the neutrino or
anti-neutrino with oscillation frequency $\omega$ and velocity $\hv$,
$\Gf$ is the Fermi coupling constant, and $n_{\omega,\hv}$ is
the number density of the corresponding neutrino. For vacuum
oscillations $\bfH_0$ is a unit vector tilted away from the unit basis
vector $\bfe_3$ in flavor space by angle $2\theta$, where $\theta$ is
the vacuum mixing angle. When a large matter density is present, one
can take into account the matter effect by setting $\bfH_0\approx\bfe_3$ and
$\theta\ll1$ \cite{Duan:2005cp,Hannestad:2006nj}. Here we assume that the
latter case is true.

Note that the coupling coefficient between two neutrinos of velocities $\hv$ and
$\hv'$ is proportional to $1-\hv\cdot\hv'$, which is because of the
current-current nature of (low-energy) weak interaction. 

Throughout this
paper we use the convention that $\hbar=c=1$, and we assume 
neutrinos to be relativistic, i.e.\ $|\hv|=1$.

In this section we focus on 
single-energy, symmetric, bipolar systems which were the first 
systems studied for collective neutrino oscillations
\cite{Kostelecky:1993yt, Kostelecky:1993dm, Pastor:2001iu,
  Duan:2005cp, Hannestad:2006nj, Raffelt:2007yz, Duan:2007mv}. 
We consider a homogeneous neutrino gas which consists of neutrinos and
anti-neutrinos of the same energy $E_0$.
We assume that  the neutrino and anti-neutrino of
the same velocity $\hv$ have equal number fluxes
(i.e.\ $n_{\omega_0,\hv}=n_{-\omega_0,\hv}$, and thus ``symmetric'' ) and opposite
flavor isospins
($\bfs_{\omega_0,\hv}(0)\approx-\bfs_{-\omega_0,\hv}(0)$, and thus ``bipolar'')
at time $t=0$.
In literature it was usually assumed that the neutrino number fluxes
were angle independent, i.e.
$n_{\omega_0,\hv}=n_\tot/4\pi$,
where $n_\tot$ is the total neutrino number density.
Here we allow a non-trivial angle distribution
\begin{align}
f_\hv = \frac{n_{\omega_0,\hv}}{n_\tot}.
\end{align}

For such a system we define
\begin{align}
\bfd_\hv &= \bfs_{\omega_0,\hv} + \bfs_{-\omega_0,\hv}, &
\bfg_\hv &= \bfs_{\omega_0,\hv} - \bfs_{-\omega_0,\hv}.
\end{align}
From Eq.~\eqref{eq:eom-s} it is easy to show that $\bfd_\hv$ and
$\bfg_\hv$ obey EoM
\begin{subequations}
\label{eq:eom-dg}
\begin{align}
\dot\bfd_\hv &= \eta\bfg_\hv\times\bfH_0 - \mu
\bfd_\hv\times\int(1-\hv\cdot\hv')f_{\hv'}\bfd_{\hv'}
\,\rmd\Omega_{\hv'},\\
\dot\bfg_\hv &= \eta\bfd_\hv\times\bfH_0 -\mu
\bfg_\hv\times\int(1-\hv\cdot\hv')f_{\hv'}\bfd_{\hv'}
\,\rmd\Omega_{\hv'}.
\end{align}
\end{subequations}
In the above equations, the dot (``$\cdot$'') symbol denotes the differentiation
with respect to dimensionless time $\tau=|\omega_0| t$,
$\eta$ is the
signature of the neutrino mass hierarchy and
$\eta=+1$ for the normal neutrino mass hierarchy (NH, $\Delta m^2 >0)$ and $-1$
for the inverted neutrino mass hierarchy (IH, $\Delta m^2<0$),
$\mu = 2\sqrt{2}\Gf |\omega_0|^{-1} n_\tot$
is a dimensionless measure of the interaction strength between
flavor isospins (as well as a measure of the neutrino density)
\footnote{The dimensionless interaction strength
$\mu$ is always positive for the examples presented in this paper. In general
$\mu$ can  be  positive or negative depending on whether the number
density of $\nu_e$ is larger or smaller than that of the other neutrino
flavor. The systems with $\mu<0$ in certain neutrino mass
hierarchy behave like those with $\mu>0$ but in the opposite mass
hierarchy. This can be seen, e.g.\ by comparing the EoM of the plus
and minus modes in Eq.~\eqref{eq:eom-twobeams}.},
and $\rmd\Omega_{\hv'}$ is the differential solid angle with respect
to direction $\hv'$. 

We assume that at $\tau=0$, all the neutrinos and anti-neutrinos
are almost purely electron-flavored, i.e
$\bfs_{\omega_0,\hv}(0)\approx-\bfs_{-\omega_0,\hv}\approx \bfe_3/2$. 
To study the development of collective oscillation modes we shall
focus on the linear regime 
in which
\begin{align}
|\bfq_\hv \equiv \bfg_\hv-\bfe_3|\ll 1.
\end{align}
In this regime Eq.~\eqref{eq:eom-dg} simplifies as
\begin{subequations}
\label{eq:eom-dq}
\begin{align}
\dot\bfd_\hv &\approx \eta\bfq_\hv\times\bfH_0 ,\\
\dot\bfq_\hv &\approx \left[\eta\bfd_\hv
+\mu\int(1-\hv\cdot\hv')f_{\hv'}\bfd_{\hv'}
\,\rmd\Omega_{\hv'}\right]\times\bfH_0,
\end{align}
\end{subequations}
where we have ignored all the terms of order $\caO(|\bfq_\hv|^2)$ or higher.

\subsection{Two Colliding Neutrino Beams}
A simple but inspiring example of bipolar system was discussed in
\cite{Raffelt:2013isa} which
consists of only two neutrino beams with angle distribution
\begin{align}
f_\hv = \frac{1}{2}\left[\delta(\varphi) +
  \delta(\varphi-\pi)\right]
\delta(\cos\vartheta),
\end{align}
where $\vartheta$ and $\varphi$ are the polar (or zenith) and
azimuthal angles of $\hv$, respectively
\footnote{The discussions here also apply to the system where two
  beams do not collide head to head if one makes the replacement 
  $\mu\rightarrow\tmu=[1-\cos(\varphi_1-\varphi_2)]\mu/2$, where
  $\varphi_{1(2)}$ is the azimuthal angle of the corresponding beam.}.

For the two-beam system we define
\begin{align}
\btd_\pm &= \frac{\bfd_1\pm\bfd_2}{2}, &
\btq_\pm &= \frac{\bfq_1\pm\bfq_2}{2},
\end{align}
where subscripts 1 and 2 refer to the beams with $\varphi=0$ and
$\pi$, respectively. From Eq.~\eqref{eq:eom-dq} we obtain
\begin{subequations}
\label{eq:eom-twobeams}
\begin{align}
\dot\btd_+ &\approx \eta \btq_+\times\bfH_0 , &
\dot\btq_+ &\approx (\eta+\mu) \btd_+\times\bfH_0, \\
\dot\btd_- &\approx \eta \btq_-\times\bfH_0,&
\dot\btq_- &\approx (\eta-\mu) \btd_-\times\bfH_0 .
\end{align}
\end{subequations}
In other words, the plus and minus modes are
independent of each other in the linear regime.

We note that in the linear regime $\bfd$'s and $\bfq$'s are
perpendicular to $\bfH_0$.
Therefore,
\begin{align}
\ddot{\tilde q}_+ &\approx -\eta(\eta+\mu) \tilde{q}_+,
\end{align}
where $\tilde q_+$ is the amplitude of $\btq_+$.
In the NH case ($\eta=+1$)
\begin{align}
\tilde q_+(\tau) \propto \cos(\gamma \tau),
\end{align}
where $\gamma = \sqrt{1+\eta\mu}$, and the plus mode is always
stable. 
In the IH case ($\eta=-1$), however,
\begin{align}
\tilde q_+(\tau) \propto \left\{\begin{array}{ll}
e^{\kappa \tau} & \text{ if } \mu > 1, \\
\cos(\gamma \tau) & \text{ if } \mu < 1,
\end{array}\right.
\end{align}
where $\kappa=\sqrt{\mu-1}$. Therefore, the plus mode is unstable only
in the IH case and when $\mu > 1,$ and it is stable otherwise.

For the minus mode we note that $(-\btd_-, \btq_-$) follow the same
EoM of $(\btd_+, \btq_+)$ but with
$\eta\rightarrow-\eta$. Therefore, the minus mode is unstable only in
the NH case and when $\mu>1$.

\begin{figure*}[t]
$\begin{array}{@{}ccc@{}}
\includegraphics[angle=0,scale=0.24]{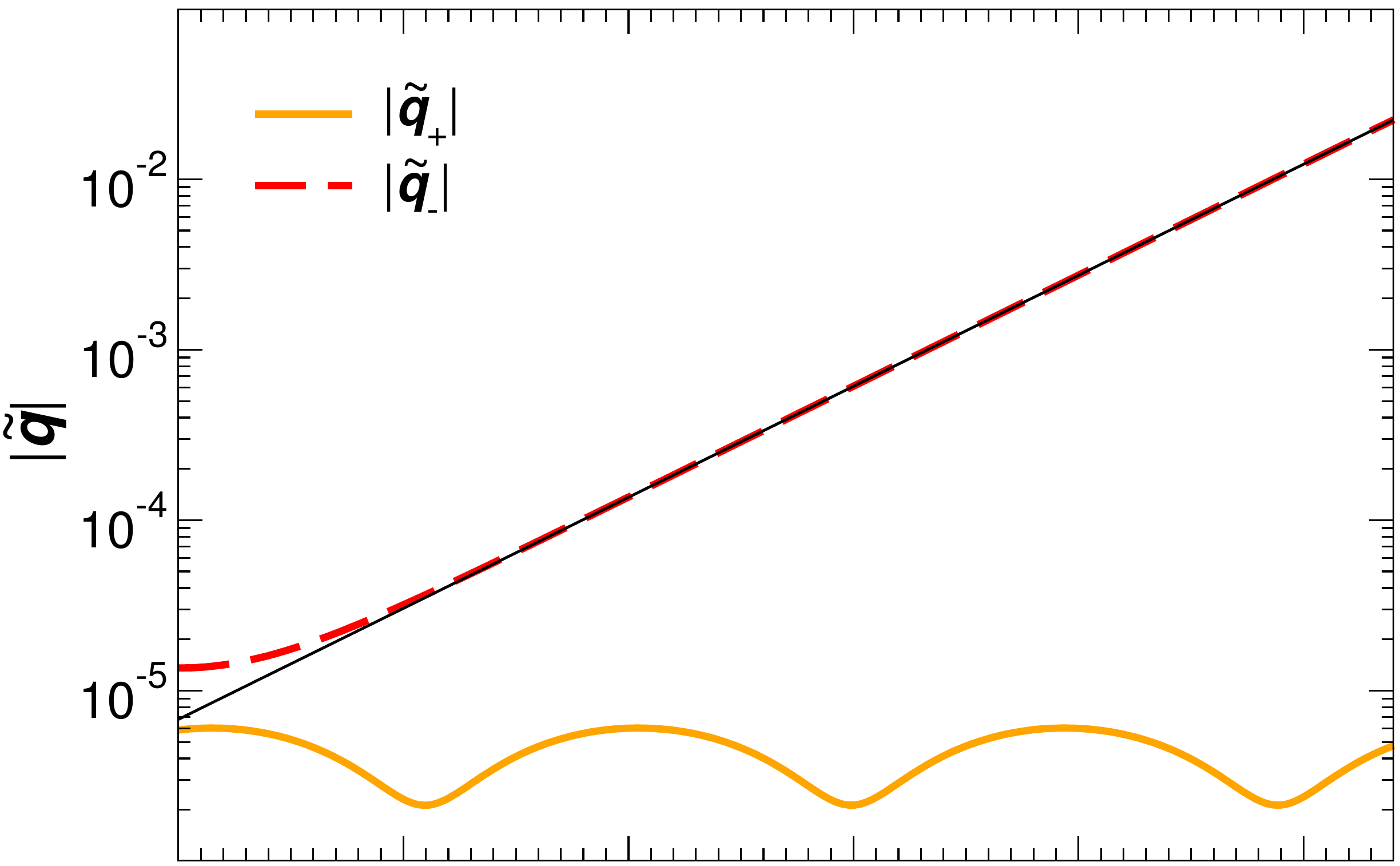} &
\includegraphics[angle=0,scale=0.24]{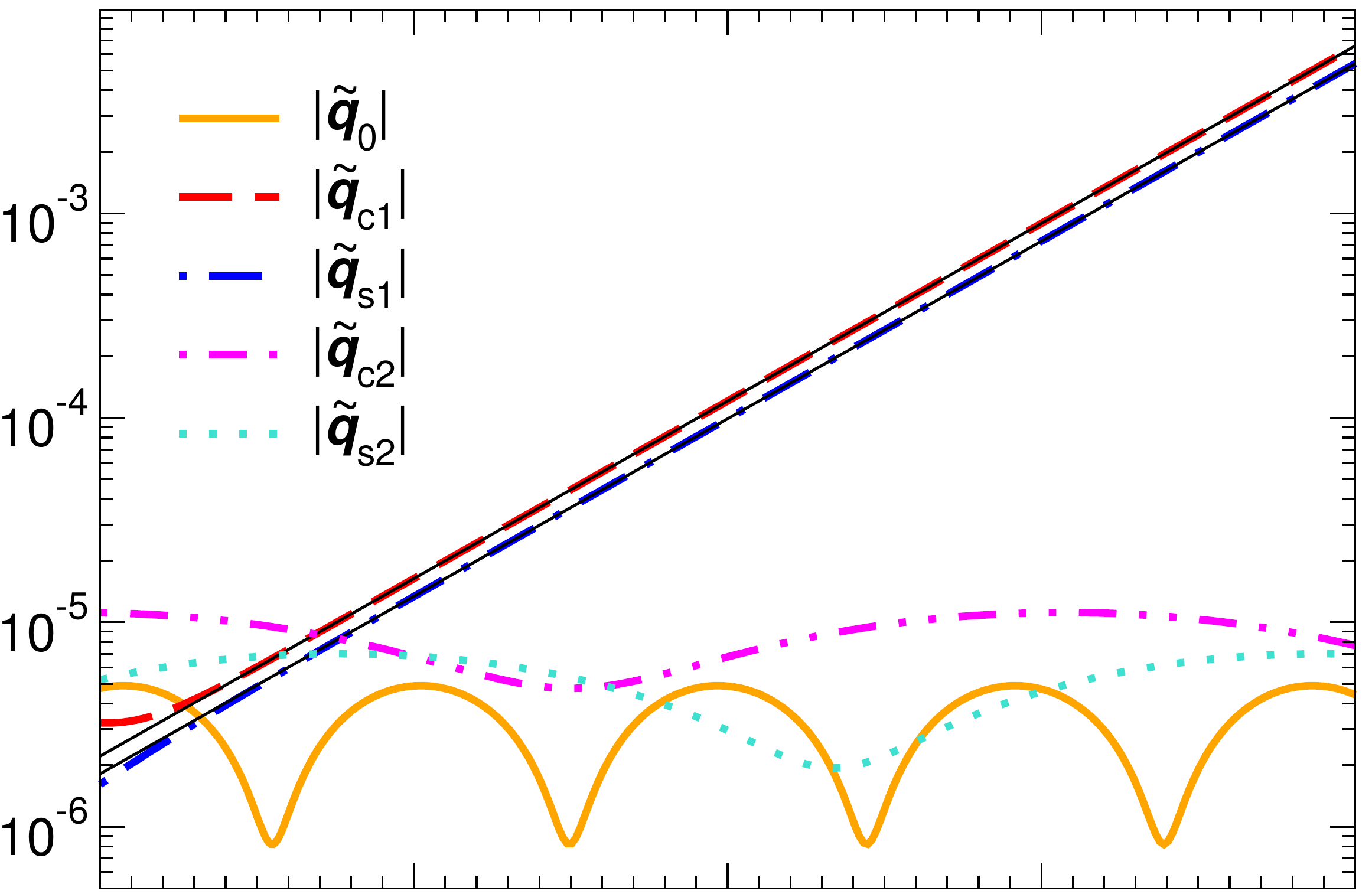} &
\includegraphics[angle=0,scale=0.24]{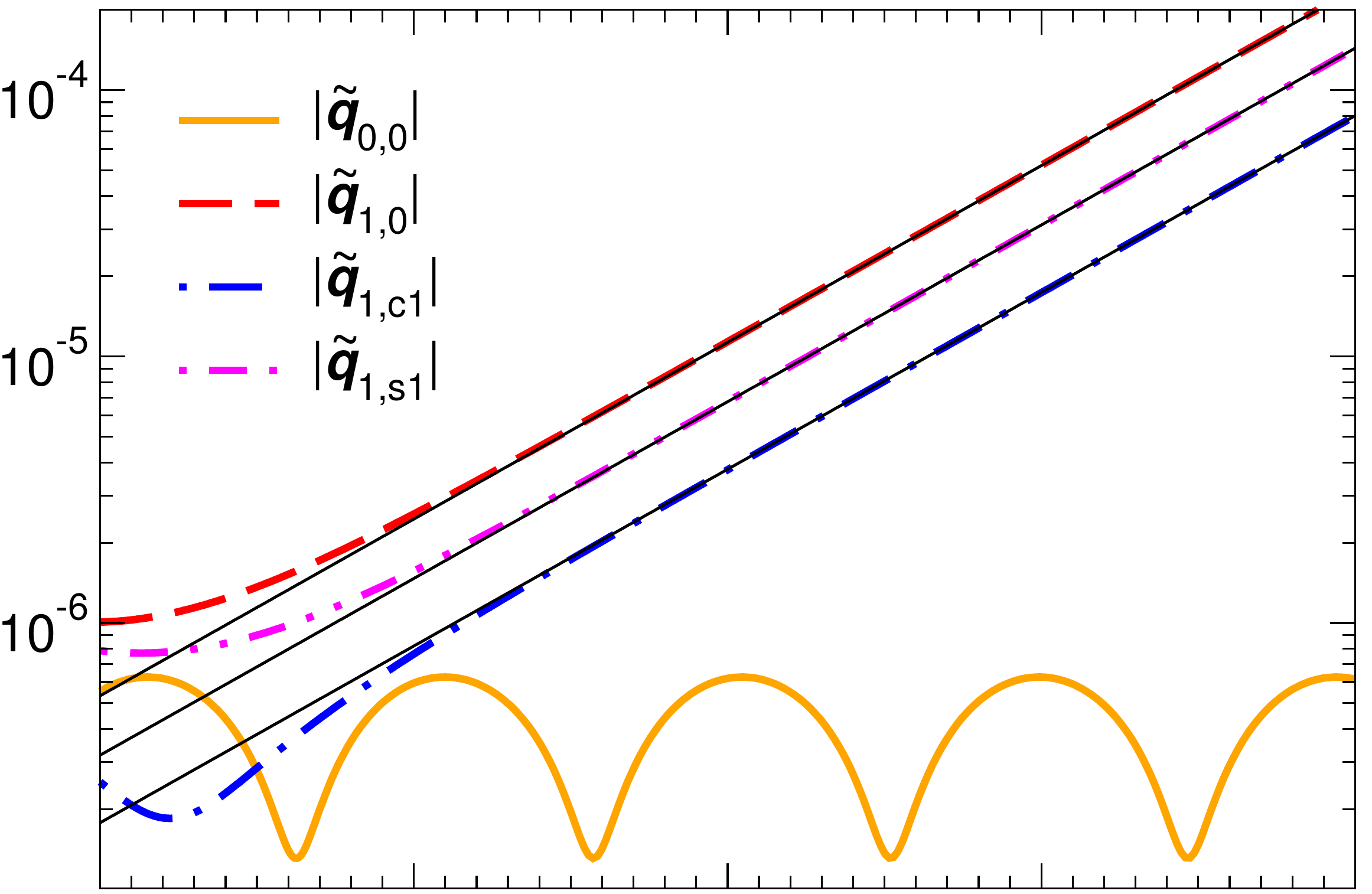}  \\
\includegraphics[angle=0,scale=0.24]{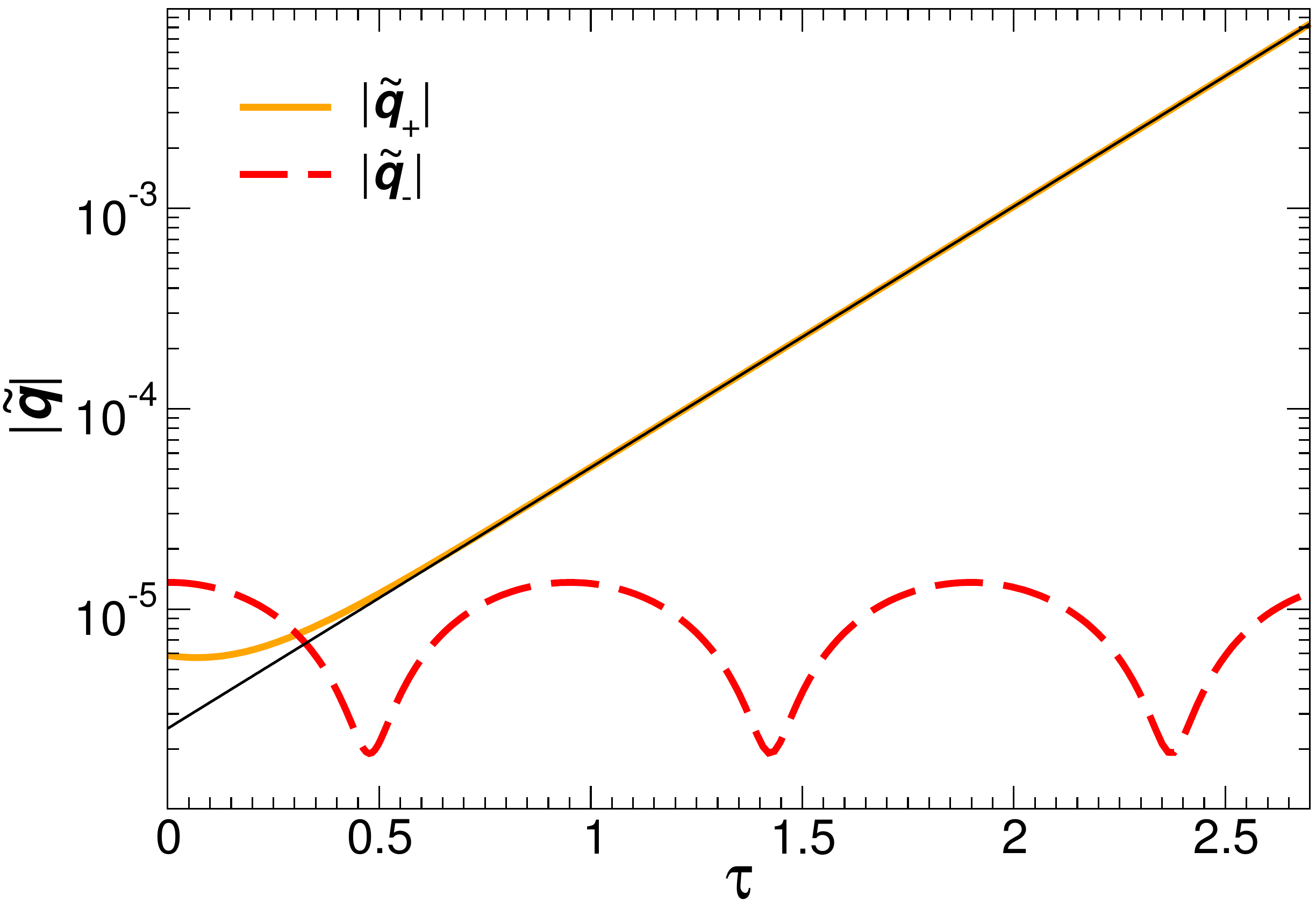} &
\includegraphics[angle=0,scale=0.24]{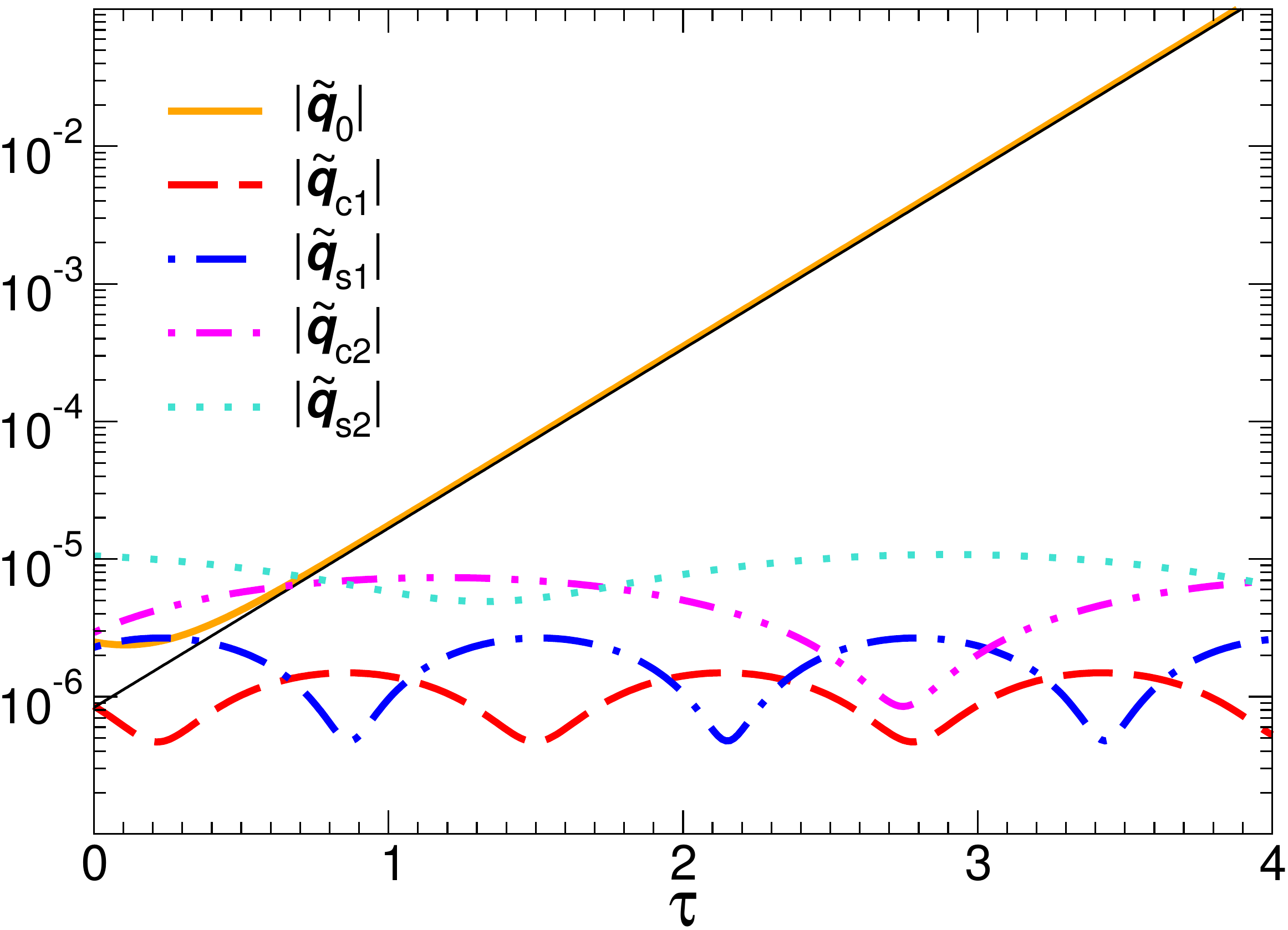} &
\includegraphics[angle=0,scale=0.24]{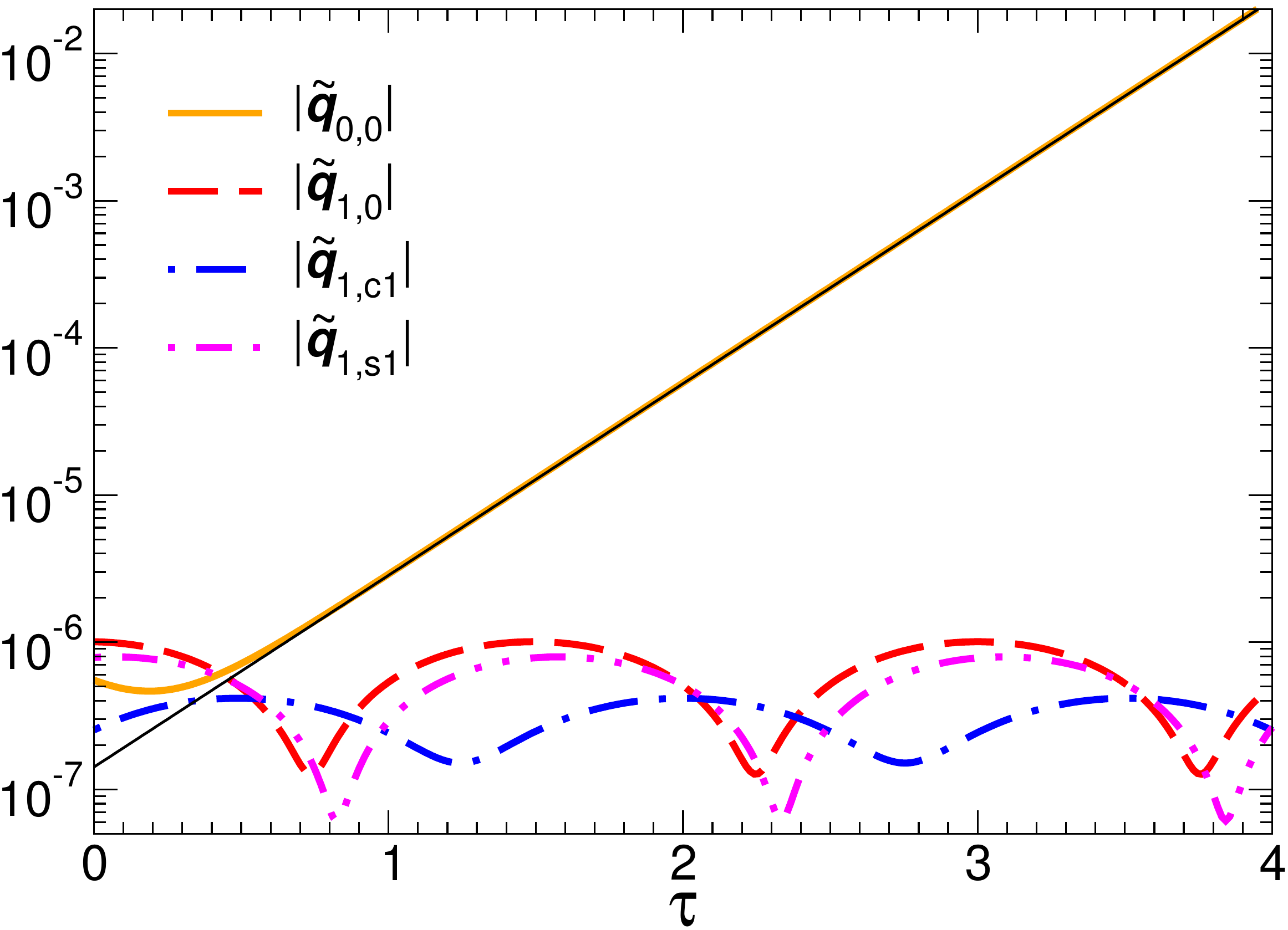}
\end{array}$
\caption{(Color online) Normal oscillation modes of monochromatic, symmetric,
  bipolar systems in the linear regime for the normal (upper panels)
  and inverted (lower panels) neutrino mass hierarchies, respectively.
  The thick lines (with legends) are the results obtained by solving the
  equation of motion, i.e.\ Eq.~\eqref{eq:eom-s}, numerically,
  and the solid thin lines have the exponential growth rates 
  $\kappa=\sqrt{\mu/2-1}$ (the middle top panel), $\sqrt{\mu/3-1}$ (the
  right top panel) 
  and $\sqrt{\mu-1}$ (the rest of the panels) with $\mu=10$.
  For numerical calculations, small random perturbations are seeded at
  $\tau=0$ so that $\bfq_\hv(0)$   are not identical for different
  $\hv$'s. 
  Left panels: The magnitudes of $\btq_\pm$ in the two-beam system.
  Middle panels: The magnitudes of $\btq_0$,
  $\btq_{\rmc m}=(\btq_m+\btq_{-m})/2$ and
  $\btq_{\rms m}=(\btq_m-\btq_{-m})/2\rmi$ in the axial-ring system.
  Right panels: The magnitudes of $\btq_{l,0}$,
  $\btq_{l, \rmc m}=(\btq_{l,m}+\btq_{l,-m})/2$ and
  $\btq_{l, \rms m}=(\btq_{l,m}-\btq_{l,-m})/2\rmi$ 
  in the isotropic-flux system.}
  \label{fig:q-tau}
\end{figure*}

In Fig.~\ref{fig:q-tau} we show the development of the plus and minus
modes in the linear regime for a two-beam system with $\mu=10$. The
numerical calculations are in good agreement with the analytic
expectations for these modes. 

\subsection{An Axial Ring of Neutrino Beams}
As another example we consider an axially symmetric ring of neutrino
beams with angle distribution
\begin{align}
f_\hv = \frac{1}{2\pi}\, \delta(\cos\vartheta).
\end{align}
For this system we define
\begin{subequations}
\begin{align}
\btd_m &= \int_0^{2\pi}\Phi_m^*(\varphi)\bfd_\varphi\,\rmd\varphi, 
\\
\btq_m &= \int_0^{2\pi}\Phi_m^*(\varphi)\bfq_\varphi\,\rmd\varphi, 
\end{align}
\end{subequations}
where
$\Phi_m(\varphi) = e^{\rmi m\varphi}/\sqrt{2\pi}$
$(m=0,\pm1, \pm2,\ldots)$
form a complete, orthonormal basis for functions of $\varphi$ defined
on $[0,2\pi)$.

Because
\begin{align}
1-\cos(\varphi-\varphi') &= 2\pi\Big[\Phi_0(\varphi)\Phi_0^*(\varphi')
\nonumber\\
&\qquad-\frac{1}{2}\sum_{m=\pm1}\Phi_m(\varphi)\Phi_m^*(\varphi')\Big],
\end{align}
the EoM of the modes with different $m$ values are decoupled in the linear
regime:
\begin{align}
\dot{\btd}_m &\approx \eta \btq_m\times\bfH_0, &
\dot{\btq}_m &\approx (\eta + \tmu_{m}) \btd_m\times\bfH_0,
\end{align}
where the effective coupling coefficient 
\begin{align}
\tmu_m = \left\{\begin{array}{ll}
\mu & \text{if } m=0,\\
-\mu/2 & \text{if } m =\pm1,\\
0 & \text{otherwise}.
\end{array}\right.
\end{align}

We note that the oscillations of the $|m|>1$ modes do not depend on
neutrino densities, and they are always stable.
We also note that the angle-independent mode (with $m=0$) behaves like the plus
mode in the two-beam system, and the $m=\pm1$ modes  behave like
the minus mode but with the effective neutrino number density reduced
by a factor of $1/2$.

These results again
agree with our numerical calculations (see Fig.~\ref{fig:q-tau}).

\subsection{Isotropic Neutrino Fluxes}
As the last example we look at the bipolar system with isotropic
neutrino number fluxes with
\begin{align} 
f_\hv = \frac{1}{4\pi}.
\end{align}
In literature it was usually assumed that the flavor content of the
neutrino fluxes was also isotropic. 
(In \cite{Raffelt:2007yz} it was assumed that the flavor content of
neutrinos was axially symmetric.)
We do not make such assumptions here.
For the isotropic-flux system we define
\begin{subequations}
\begin{align}
\btd_{l,m} &= \int Y_{l,m}^*(\hv)\bfd_\hv\,\rmd\Omega_\hv, 
\\
\btq_{l,m} &= \int Y_{l,m}^*(\hv)\bfq_\hv\,\rmd\Omega_\hv, 
\end{align}
\end{subequations}
where $Y_{l,m}$ are the spherical harmonics. 
Because
\begin{align}
1-\hv\cdot\hv' &= 4\pi\Big[Y_{0,0}(\hv)Y_{0,0}^*(\hv')
\nonumber\\
&\qquad-\frac{1}{3}\sum_{m=0,\pm1} Y_{1,m}(\hv) Y_{1,m}^*(\hv')\Big],
\end{align}
the EoM of the modes with different $(l,m)$ values are decoupled in the linear
regime:
\begin{align}
\dot{\btd}_{l,m} &\approx \eta \btq_{l,m}\times\bfH_0, &
\dot{\btq}_{l,m} &\approx (\eta + \tmu_l) \btd_{l,m}\times\bfH_0,
\end{align}
where the effective coupling coefficient 
\begin{align}
\tmu_l = \left\{\begin{array}{ll}
\mu & \text{if } l=0,\\
-\mu/3 & \text{if } l =1,\\
0 & \text{otherwise}.
\end{array}\right.
\end{align}

We note that the $l>1$ modes in the isotropic-flux system are like
the $|m|>1$ modes in the axial-ring system, and they are always stable.
We also note that the monopole mode ($l=0$) behaves like the plus
mode in the two-beam system, and that the dipole modes ($l=1$) behave like
the minus mode but with the effective neutrino number density reduced
by $2/3$.

These results indeed
agree with our numerical calculations (see
Fig.~\ref{fig:q-tau}).

\section{Extensions and Discussion%
\label{sec:discussions}}

The techniques discussed in Section~\ref{sec:bipolar} can be applied to 
more general systems. A simple but useful generalization is the
monochromatic, homogeneous, isotropic (in terms of neutrino number fluxes),
non-symmetric bipolar gas. In a non-symmetric system the number
densities of the neutrino 
and anti-neutrino are not the same but are related by
\begin{align}
\bar{n}_\tot = \alpha n_\tot,
\end{align}
where $\bar{n}_\tot$ is the total number density of the anti-neutrino.
Again we assume that
$\bfs_{\omega_0,\hv}\approx\bfs_{-\omega_0,\hv}\approx \bfe_3/2$ at $\tau=0$.
For such a system we define
\begin{subequations}
\begin{align}
\bfd_\hv &= \bfs_{\omega_0,\hv} + \alpha\bfs_{-\omega_0,\hv} -
\frac{(1-\alpha)}{2} \,\bfe_3, \\
\bfq_\hv &= \bfs_{\omega_0,\hv} - \alpha\bfs_{-\omega_0,\hv} -
\frac{(1+\alpha)}{2} \,\bfe_3.
\end{align}
\end{subequations}
In the linear regime where $|\bfq_\hv|\ll1$ we obtain
\begin{subequations}
\begin{align}
\dot\bfd_\hv &\approx \left[\eta\bfq_\hv 
+ (1-\alpha)\frac{\mu}{2}
\int(1-\hv\cdot\hv')f_{\hv'}\bfd_{\hv'}\,\rmd\Omega_{\hv'}\right] 
\times\bfH_0 
\nonumber\\
&\quad- (1-\alpha)\frac{\mu}{2} \bfd_\hv\times\bfH_0,\\
\dot\bfq_\hv &\approx 
\left[\eta\bfd_\hv +
(1+\alpha)\frac{\mu}{2}
\int(1-\hv\cdot\hv')f_{\hv'}\bfd_{\hv'}\,\rmd\Omega_{\hv'}\right]
\times\bfH_0 
\nonumber\\
&\quad- (1-\alpha)\frac{\mu}{2} \bfq_\hv\times\bfH_0.
\end{align}
\end{subequations}
In the reference frame which rotates about $\bfH_0$ by frequency
$(1-\alpha)\mu/2$ we obtain
\begin{subequations}
\begin{align}
\dot\btd_{l,m}&\approx\left[\eta\btq_{l,m} 
+ (1-\alpha)\frac{\tmu_l}{2}\,\btd_{l,m}\right]\times\bfH_0, \\
\dot\btq_{l,m}&\approx\left[\eta+(1+\alpha)\frac{\tmu_l}{2}\right]
\,\btd_{l,m}\times\bfH_0.
\end{align}
\end{subequations}
The EoM of $\bfd$'s and $\bfq$'s again decouple in the spherical
basis. 

We note that the $l>1$ modes oscillate with frequencies $\omega$ in
the co-rotating frame and that they are always stable. 
The monopole mode (with $l=0$ and $\tmu_0=\mu)$ has been well
studied in literature 
\cite{Duan:2005cp,Hannestad:2006nj,Duan:2007mv,Banerjee:2011fj}. 
This mode is unstable only in the IH case and when
\begin{align}
\frac{4}{(1+\sqrt{\alpha})^2} < \mu < \frac{4}{(1-\sqrt{\alpha})^2}.
\label{eq:mu-bipolar}
\end{align}
For the dipole modes (with $l=1$ and $\tmu_1=-\mu/3$) we note that
$(-\bfd_{1,m},\bfq_{1,m})$ obey the same EoM of
$(\bfd_{0,0},\bfq_{0,0})$ except with replacements
\begin{align}
\eta&\rightarrow-\eta, &
\mu &\rightarrow \frac{\alpha\mu}{3}, &
\alpha&\rightarrow \alpha^{-1}.
\label{eq:replacements}
\end{align}
This implies that the dipole modes are unstable only in the NH case
and when
\begin{align}
 \frac{12}{(1+\sqrt{\alpha})^2} < \mu < \frac{12}{(1-\sqrt{\alpha})^2}.
\label{eq:mu-maa}
\end{align}

One can further generalize these techniques to homogeneous and
isotropic neutrino gases with continuous energy spectra. In such
systems the oscillation modes with different $(l,m)$ are not coupled
in the linear regime. The monopole modes with different $\omega$'s are
coupled, and these modes can engender the angle-independent collective
flavor transformation which was studied in the literature for neutrino
oscillations in the early Universe and that for the single-angle
approximation of supernova neutrinos. The dipole modes with different
$\omega$'s are also coupled. The results of the monopole modes can be
applied to the dipole modes with appropriate replacements similar to
Eq.~\eqref{eq:replacements}. 

We note that, however, the multipole modes are no longer decoupled
once the unstable monopole or dipole modes grow out of the linear
regime. The convolution of the multipole modes can but not always result in
kinematic decoherence
\cite{Raffelt:2007yz,EstebanPretel:2007ec}. In any case, it is not
always justified to assume that the flavor content of neutrino fluxes
will remain isotropic even if both the number fluxes and the flavor content of a
homogeneous neutrino gas are (approximately) isotropic initially.

The geometric nature of the supernova environment is
much more complicated than that of a homogeneous neutrino
gas. A generalization to the ``neutrino bulb model''
\cite{Duan:2006an} was studied in \cite{Raffelt:2013rqa,
  Mirizzi:2013rla, Mirizzi:2013wda}. Like in the original  bulb
model, both the number fluxes and the flavor content of neutrinos 
 in the generalized bulb model are
spherically symmetric about the center of the neutron star at all
radii, and they are also axially symmetric
about any radial 
direction on the surface of the neutron star. In the generalized bulb model,
however, the flavor content of neutrinos are not assumed to be axially
symmetric at all radii.

Because of the initial axial symmetry on the surface of the neutron
star, one can apply the techniques discussed for the
axial-ring and  isotropic-flux models in Section~\ref{sec:bipolar} to
the generalized bulb model.
It is straightforward to see that the $|m|>1$ modes should always be stable in
the linear regime. It is also obvious that 
the newly discovered MAA modes have $m=\pm1$
and behave like the minus modes in the two-beam systems.
The well known bipolar/bimodal mode and the 
newly discovered multi-zenith-angle (or MZA) mode
\cite{Raffelt:2013rqa} both have $m=0$ 
\footnote{In a private communication with the author 
G.~Raffelt pointed out that,
in contrast to the three degenerate dipole modes in a homogeneous,
isotropic neutrino gas, the MZA mode and two MAA modes of supernova
neutrinos are not degenerate 
because of the special geometry of the neutrino bulb model.}.

As in the axial-ring model, the bipolar and MAA modes in the
generalized bulb model are
responsible for the flavor instabilities in the opposite neutrino mass
hierarchies. This result indeed agrees with what was shown in
\cite{Raffelt:2013rqa}. In addition, the effective neutrino number
densities of the MAA modes is a fraction of that of the bipolar
mode. (This fraction can be different from those in the axial-ring and
isotropic-flux  models  because of their different dependence of the
neutrino beams and, therefore, the neutrino-neutrino scattering
potential, on polar angles.) 
This implies that the MAA modes can become unstable closer to the
neutron star than the bipolar mode does. 
[One can see this by, e.g.\ comparing Eq.~\eqref{eq:mu-maa} to
  \eqref{eq:mu-bipolar}.] 
This result also agrees with what was shown in \cite{Raffelt:2013rqa}. 

We emphasize that the generalized bulb model studied in
\cite{Raffelt:2013rqa, Mirizzi:2013rla, Mirizzi:2013wda} is not
entirely self-consistent because the MAA modes break the axial
symmetry about the radial direction. Even if the bipolar mode is
unstable in the linear regime, the
azimuthal-angle dependent and independent modes become coupled once
the unstable mode(s) grows out of the linear regime, which can also break
the axial symmetry. In a supernova neutrino model
the spherical symmetry breaks down once the axial symmetry about any radial
direction is lost. Therefore, one needs to go beyond the spherical
supernova model in studying azimuthal-angle-dependent collective
neutrino oscillations even if everything else (the matter
density, over all neutrino number fluxes, etc) in the supernova is
spherically symmetric. 

\section{Conclusions%
\label{sec:conclusions}}

We have studied the flavor oscillation modes in homogeneous neutrino
gases. We have identified the normal modes of neutrino oscillations in the
linear regime for the
monochromatic two-beam, axial-ring and isotropic-flux neutrino systems
with the symmetric bipolar configuration.
We have shown that, because of the current-current
nature of (low-energy) weak interaction, only the monopole and dipole
modes (or the plus/symmetric and minus/anti-symmetric modes in the case of a
two-beam system) could trigger significant flavor transformation in the
opposite neutrino mass hierarchies. 
All other multipole modes are stable in the linear regime.

We have also discussed how to generalize the results found in the
above models to  
other systems such as non-symmetric bipolar gases and 
homogeneous, isotropic gases with continuous neutrino energy spectra. 

Our study provides new insights into the recent discovery of
the MAA flavor instabilities of supernova neutrinos. The 
bipolar mode and the MZA mode in the generalized neutrino bulb model
are the $m=0$ modes while the MAA modes have
$m=\pm1$. The
bipolar mode and the MZA/MAA modes are responsible for flavor
instabilities in different neutrino mass hierarchies. 
The MAA modes can become unstable at a radius smaller
than that of the bipolar mode in
the opposite neutrino mass hierarchy, which makes them more interesting for
supernova nucleosysthesis. However, we also note that, if the
azimuthal-angle dependent modes are ever important (whether inside or
outside the linear regime), supernova neutrinos cannot be treated
self-consistently in a spherical supernova model anymore.

\appendix
\section{Equations of Motion for Neutrino Oscillations%
\label{sec:nfis}}
\subsection{Flavor Density Matrix}
For two-flavor neutrino oscillations the flavor density matrix of the
neutrino is 
\begin{equation}
 \rho = \begin{bmatrix}
\rho_{ee} & \rho_{ex} \\ \rho_{ex}^* & \rho_{xx} \end{bmatrix}.
\end{equation}
The diagonal elements of $\rho$ give the occupation numbers of the
neutrino in the $e$ and $x$ flavors, respectively, and
the off-diagonal elements contain the information of flavor mixing. 
The flavor density matrix $\bar\rho$ for the anti-neutrino is defined
in a similar way.
In a homogeneous neutrino gas $\rho$ and $\bar\rho$ have no
spatial dependence. In absence of collisions, they obey the equations
of motion (EoM) \cite{Sigl:1992fn}
\begin{subequations}
\begin{align}
\rmi\td\rho_\vp &=
 [\sfH_0 + \sfH_\matt +
\sfH_{\nu\nu}(t,\hv), \, \rho_\vp], \\
\rmi\td\bar\rho_\vp &=
 [\sfH_0 - \sfH_\matt -
\sfH_{\nu\nu}^*(t,\hv), \, \bar\rho_\vp]
\label{eq:eom-rho}
\end{align}
\end{subequations}
for a homogeneous neutrino gas,
where $\vp$ is the momentum of the neutrino or anti-neutrino, 
$\hv=\vp/E$ is its velocity, and $\sfH_0$,
$\sfH_\matt$ and $\sfH_{\nu\nu}$ are the vacuum, matter and neutrino
self-coupling (or neutrino-neutrino forward-scattering) parts of the
Hamiltonian, respectively. 

The vacuum Hamiltonian is
\begin{align}
\sfH_0 = \frac{\Delta m^2}{4 E} 
\begin{bmatrix} -\cos 2\theta & \sin2 \theta \\
\sin2\theta & \cos2\theta \end{bmatrix},
\end{align}
where $\theta$ is the vacuum mixing angle, and 
$\Delta m^2 = m_2^2 - m_1^2$ is the mass-squared difference between
mass eigenstates $\ket{\nu_1}$ and $\ket{\nu_2}$. 

The matter Hamiltonian is
\begin{align} 
\sfH_\matt =  \sqrt{2}\Gf n_e \begin{bmatrix} 1 &
0 \\ 0 & 0 \end{bmatrix},
\end{align}
where $\Gf$ is the Fermi coupling constant, and $n_e$ is the net electron
number density. For a homogeneous neutrino gas, a large matter density
effectively sets the vacuum mixing angle to a small value \cite{Duan:2005cp,
  Hannestad:2006nj, Duan:2008za}. We assume this is the case, and we set
 $\sfH_\matt=0$ and $\theta\approx 0$.

The neutrino self-coupling Hamiltonian is 
\cite{Fuller:1987aa, Notzold:1988kx, Pantaleone:1992xh}
\begin{align} 
\sfH_{\nu\nu}(t,\hv) = \sqrt{2}\Gf\sum_{\vp'}(1-\hv\cdot\hv') 
[\rho_{\vp'}(t) - \bar\rho_{\vp'}^*(t)].
\end{align}

\subsection{Neutrino Flavor Isospin}

To visualize the evolution of  flavor density matrices, we use
the concept of the neutrino flavor isospin 
defined in \cite{Duan:2005cp}. The flavor density matrix $\rho$ and flavor
isospin $\bfs$ of the same neutrino are related by
\begin{subequations}
\begin{align}
\rho_\vp &= \frac{n_\vp}{2} +
n_{\omega,\hv}\bsigma\cdot\bfs_{\omega,\hv}, \\
\bar\rho^*_\vp &= \frac{\bar{n}_\vp}{2} +
n_{-\omega,\hv}\bsigma\cdot(-\bfs_{-\omega,\hv}), 
\end{align}
where 
\begin{align}
n_\vp = \rho_{ee,\vp} + \rho_{xx,\vp} 
\quad\text{and}\quad
\bar{n}_\vp = \bar\rho_{ee,\vp} + \bar\rho_{xx,\vp}
\end{align}
\end{subequations}
are the number densities of the neutrino and anti-neutrino of momentum
$\vp$, respectively,
$\sigma_i$ ($i=1,2,3$) are the Pauli matrices, 
\begin{align}
\omega = \frac{\Delta m^2}{2 E}
\end{align}
is the vacuum oscillation frequency, and $n_{\pm\omega,\hv}$ is the
effective density distribution of the neutrino.
We use the convention that a bold symbol
(e.g.\ $\bfs$) denotes a vector in flavor space, and a symbol with
the vector hat (e.g.\ $\vp$) denotes a vector in coordinate space.

The flavor density matrix formalism and the flavor isospin
formalism are completely equivalent in the two-flavor
scheme. In absence of collision neither the trace of a flavor
density matrix nor the effective density distribution $n_{\omega,\hv}$
changes over time. Therefore, we  ignore the traces of flavor
density matrices which have no impact on neutrino oscillations.

Also note that the flavor content of a neutrino is represented by a flavor
isospin with $\omega > 0$ ($\omega<0$) for the normal (inverted) mass
hierarchy with $\Delta m^2 > 0$ ($\Delta m^2 <0$). If the neutrino is in a
pure (flavor) quantum state and is represented by flavor wavefunction
$\psi$, then the 
flavor isospin can be equivalently defined as
\begin{align}
\bfs &= \psi^\dagger \frac{\bsigma}{2} \psi.
\end{align}
Similarly, the flavor content
of an anti-neutrino is represented by a flavor
isospin with $\omega < 0$ ($\omega>0$) if  $\Delta m^2 > 0$ 
($\Delta m^2 <0$). For an anti-neutrino represented by flavor wavefunction
$\psi$, its flavor isospin is
\begin{align}
\bfs &= (\sigma_2 \psi)^\dagger \frac{\bsigma}{2}(\sigma_2\psi).
\end{align}
In particular, a pure $\nu_e$ is represented by $\bfs=\bfe_3/2$, and a
pure $\bar\nu_e$ is represented by $\bfs=-\bfe_3/2$, where
$\bfe_i$ ($i=1,2,3$) are the basis unit vectors in flavor space like
$\hat{x}$, $\hat{y}$ and $\hat{z}$ in coordinate space.

The EoM for the flavor isospin is
\begin{align}
\td\bfs_{\omega,\hv} &= 
 - 2\sqrt{2}\Gf \bfs_{\omega,\hv}\times \sum_{\omega',\hv'} 
(1-\hv\cdot\hv') n_{\omega',\hv'} \bfs_{\omega',\hv'}
\nonumber\\
&\qquad + \bfs_{\omega,\hv}\times\omega \bfH_0, 
\end{align}
where $\bfH_0\approx\bfe_3$ (because we set vacuum mixing angle
$\theta\approx 0$).
 
One of the main advantages of the concept of flavor isospin is that
the neutrino and the anti-neutrino are treated on an equal footing in
this formalism.

\begin{acknowledgments}
The author thanks G.\ Raffelt for reading the draft and providing
helpful suggestions. The author also thanks
Y.-Z.\ Qian and S.\ Shashank for useful discussions.
This work was supported in part by 
DOE grant DE-SC0008142 at UNM.
\end{acknowledgments}

\bibliography{ref}

\end{document}